\documentclass[11pt]{article}
\usepackage{amsmath,amssymb,amsthm}
\usepackage{booktabs}
\usepackage{array}
\usepackage{graphicx}
\usepackage{placeins}
\usepackage[round]{natbib}
\usepackage{url}

\graphicspath{{figures/}}

\newcommand{\R}{\mathbb{R}}
\newcommand{\E}{\mathrm{E}}
\newcommand{\pr}{\mathrm{pr}}
\newcommand{\len}{\mathrm{len}}
\newcommand{\op}{o_{\mathrm{p}}}

\newcommand{\alttext}[1]{}

\theoremstyle{plain}
\newtheorem{theorem}{Theorem}
\newtheorem{proposition}{Proposition}

\begin{document}

\title{Geometry of tail allocation in conformal prediction intervals}
\author{Tianying Wang\thanks{Department of Statistics, Colorado State University,
Fort Collins, Colorado, U.S.A. Email: Tianying.Wang@colostate.edu.}}
\date{}

\maketitle
\thispagestyle{empty}

\begin{abstract}
Lower and upper errors of a two-sided conformal prediction interval can have
different scientific consequences. The division of target miscoverage between
the two endpoints determines the corresponding tail-specific guarantees and
can alter interval length at first order when tail scales differ. We
characterize this allocation-length relation after separate one-sided split
calibration, which preserves the tail-specific guarantees and marginal
coverage whenever the allocation is selected independently of the calibration
sample. Tail-quantile response to proportional rescaling determines the
resulting length geometry. For
regularly varying tails, normalized length converges to
\(g_\gamma(c)=c^{-\xi}+\gamma(1-c)^{-\xi}\), where \(c\) is the upper-tail
allocation fraction, \(\xi\) is the tail index, and \(\gamma\) is the
lower-to-upper tail-scale ratio. A dominant tail produces a boundary optimum
and makes the equal-tail interval asymptotically \(2^\xi\) times as long as the
optimum. Comparable tails produce an interior optimum, with equal-tail
allocation optimal only at matching scales. An empirical allocation rule
attains the corresponding optimum without estimating tail parameters. In the
de Haan class the effect moves to an additive scale. Calibration resolution
determines whether ordinary ranks can realize these allocations. When
calibration tail counts remain bounded, two-sided rank feasibility also
constrains the allocation. Tail homogeneity transfers the length relation over
covariates,
while opposite dominant tails preclude one globally efficient allocation.
\end{abstract}

\noindent
\textbf{Keywords:} Conformal prediction; Extreme value theory; Prediction
interval; Quantile regression; Regular variation; Tail allocation.

\pagestyle{plain}

\section{Introduction}
\label{sec:introduction}

Miscoverage of a two-sided prediction interval occurs when the response falls
below its lower endpoint or exceeds its upper endpoint. These events can have
different scientific consequences. When their two probabilities are prescribed
separately, they form part of the inferential target and should remain fixed.
When only total miscoverage is prescribed, its division between the two tails
remains available. That division determines the tail-specific guarantees, the
endpoint quantile levels, and potentially the length of the final interval. We
study this latter setting.

Conformal prediction turns a fitted predictive model into a set with
finite-sample marginal coverage under exchangeability
\citep{vovk2005algorithmic,lei2018distribution}. For a real-valued response,
the usual connected interval assigns half of the target miscoverage to each
tail, as in the standard form of conformalized quantile regression
\citep{romano2019conformalized}. Marginal validity, however, requires only that
the two tail-specific levels sum to the target level. If the upper tail is
heavy while the lower quantile changes on a smaller scale, assigning more
miscoverage to the upper tail can substantially reduce the upper endpoint. If
both tails are heavy, the same allocation can enlarge the interval through the
lower endpoint. Equal-tail allocation is therefore not generally neutral, and
separate conformal calibration need not remove its length consequence.

Population shortest intervals allocate excluded probability by balancing
endpoint densities \citep{liu2015spin,brehmer2021intervals}, while one-sided
conformal constructions provide tail-specific guarantees
\citep{cuonzo2026tails}. At a fixed target level, set-selection, histogram,
distributional, and direct interval procedures shorten or reshape conformal
sets \citep{yang2025selection,kiyani2024length,lebars2025volume,
sesia2021histograms,chernozhukov2021distributional,izbicki2022cdsplit,pouplin2024relaxed,
sesia2020comparison}. Interquantile and skew-adaptive proposals alter fitted
ranges, interval centers, or calibration scores
\citep{luo2025thresholded,guo2026interquantile,marques2026skew,
su2026cooptimization,zou2026percentile}, and localized conformal methods vary
calibration across covariate neighborhoods \citep{guan2023localized}. These
lines provide the neighboring ingredients but leave a scaling question open.
After two fitted endpoints have been calibrated separately, how does a
proportional redistribution of a shrinking miscoverage level change the final
interval length?

Two quantities organize the answer. The first is the response of an endpoint
quantile when its tail probability is multiplied by a fixed fraction. Regular
variation produces a multiplicative response on the leading tail-quantile
scale, whereas de Haan variation produces an additive response on a smaller
auxiliary scale \citep{bingham1987regular,dehaan2006extreme}. Within the
regularly varying class, the relative sizes of the two endpoint scales place
the optimum at the boundary or in the interior. The second quantity is the
number of calibration observations in the target tails. Diverging tail counts
smooth the conformal ranks, while bounded tail counts make rank discreteness
visible at first order.

Our starting point is an exact max-min identity that places fitted tail
quantiles and calibration order statistics on a common scale. Under regularly
varying tails, the resulting first-order allocation-length function is
explicit. If \(c\) is the upper-tail allocation fraction and \(\gamma\) is the
lower-to-upper tail-scale ratio, length normalized by the upper
\((1-\alpha)\)-quantile converges to
\[
g_\gamma(c)=c^{-\xi}+\gamma(1-c)^{-\xi}.
\]
When one tail dominates, the optimum is at the boundary and the equal-tail
interval is asymptotically \(2^\xi\) times as long as the optimum. When both
tails contribute at first order, the optimum is interior, with equal-tail
allocation optimal only when the two tail scales match. The same function
therefore gives both the equal-tail penalty and the unique optimal allocation.
An empirical allocation rule that minimizes fitted width over a candidate
allocation grid spanning the finite-rank allocation domain attains this optimum
without estimating \(\xi\), \(\gamma\), or first identifying the regularly
varying branch. Because the allocation is selected without calibration
observations, separate one-sided calibration preserves the tail-specific
guarantees and finite-sample marginal coverage.

The two organizing quantities also describe where this conclusion changes
form. In the de Haan class, allocation leaves first-order relative length
unchanged but produces an additive contrast. When calibration tail counts
remain bounded, a random finite-rank limit replaces the smooth length relation,
and a finite two-sided interval becomes possible only after the limiting total
tail count exceeds two. Under tail homogeneity, the regularly varying relation
transfers uniformly over covariates, whereas groups with opposite dominant
tails preclude one globally efficient allocation.

The endpoint allocations \(c=0\) and \(c=1\) reduce to lower and upper
one-sided split calibration, and intersecting the resulting sets gives a
tail-specific two-sided interval \citep{cuonzo2026tails}. Censoring-specific
lower survival bounds pursue the one-sided objective under a different
observation structure \citep{candes2023survival}. Beyond ordinary conformal rank
resolution, extreme conformal prediction uses tail extrapolation
\citep{pasche2026extreme}. Our analysis contains the fully observed endpoint
constructions, studies the interior allocation fractions, and identifies the
finite-rank limit from the ordinary-rank side. It therefore connects
tail-specific validity to the calibrated length geometry that remains visible
before extrapolation is required.

\section{Tail-allocated conformal intervals}
\label{sec:method}

\subsection{Construction and finite-sample validity}
\label{sec:construction}

Let \((X_i,Y_i)\) be exchangeable observations with \(Y_i\in\R\). The data are
divided into fitting, selection, and calibration samples. The calibration
sample has size \(m\). The fitting sample produces conditional quantile
estimates \(\widehat q_p(x)\) over the probability levels considered below.
The selection sample is used only to compare fitted widths. It may be omitted
in the scalar empirical-quantile setting, where the fitted endpoints themselves
define the comparison. A fitted endpoint or allocation is called
calibration-independent if it is measurable with respect to data independent
of the calibration sample and the test pair.

Write \(\alpha_L+\alpha_U=\alpha\) for the lower and upper tail-specific
miscoverage levels, and
let \(c=\alpha_U/\alpha\) be the upper-tail allocation fraction. For a
candidate \(c\), the fitted core interval is
\(\widehat I_c(x)=[\widehat q_{(1-c)\alpha}(x),
\widehat q_{1-c\alpha}(x)]\).
If the selection sample has size \(n_{\rm se}\), with covariates
\(X^{\rm se}_1,\ldots,X^{\rm se}_{n_{\rm se}}\), define
\[
T_{n_{\rm se}}(c)=n_{\rm se}^{-1}\sum_{j=1}^{n_{\rm se}}
\{\widehat q_{1-c\alpha}(X^{\rm se}_j)
-\widehat q_{(1-c)\alpha}(X^{\rm se}_j)\}.
\]
Set \(\mathcal C_m(\alpha)=[b_m/\alpha,1-b_m/\alpha]\), where
\(b_m=1/(m+1)\). This finite-rank allocation domain is nonempty exactly when
\(\alpha\ge2b_m\). If
\(\alpha<2b_m\), no allocation gives both endpoints a finite one-sided
calibration rank. When the set is nonempty, let \(\mathcal G_m\) be a finite
subset of \(\mathcal C_m(\alpha)\) that contains both endpoints and whose
largest gap tends to zero along the asymptotic sequence. Choose the smallest
minimizer \(\widehat c\) of \(T_{n_{\rm se}}(c)\) over \(\mathcal G_m\).
The value \(b_m\) is the smallest tail probability for which a one-sided
calibration rank is finite. In the scalar setting, \(T_{n_{\rm se}}(c)\) is replaced by
the width between empirical fitting-sample quantiles. The equal-tail and rank-floor
choices are \(c=1/2\) and \(c=1-b_m/\alpha\), respectively. The former treats
the tails symmetrically. The latter uses the smallest finite lower-tail
miscoverage level and assigns the remainder to the upper tail.

For a selected allocation, define upper and lower calibration scores
\(S_i^U=Y_i^{\rm cal}-\widehat q_{1-\alpha_U}(X_i^{\rm cal})\) and
\(S_i^L=\widehat q_{\alpha_L}(X_i^{\rm cal})-Y_i^{\rm cal}\).
Let \(S^U_{(j)}\) and \(S^L_{(j)}\) denote the ordered scores, and set
\(k_U=\lceil(m+1)(1-\alpha_U)\rceil\) and
\(k_L=\lceil(m+1)(1-\alpha_L)\rceil\).
The nonnegative corrections are
\(R^U=\max(S^U_{(k_U)},0)\) and
\(R^L=\max(S^L_{(k_L)},0)\). If a requested rank exceeds \(m\), the
corresponding correction is infinite. Write \(R_s^U\) and \(R_s^L\) for the
corrections obtained when the corresponding tail-specific miscoverage level is
\(s\). Define
the calibrated endpoint functions
\(\widehat U_s(x)=\widehat q_{1-s}(x)+R^U_s\) and
\(\widehat L_s(x)=\widehat q_s(x)-R^L_s\).
The interval for a new covariate value \(x\) is
\begin{equation}
\label{eq:interval}
\widehat C_{\alpha_L,\alpha_U}(x)=
[\widehat q_{\alpha_L}(x)-R^L,
\widehat q_{1-\alpha_U}(x)+R^U].
\end{equation}

\begin{proposition}
\label{prop:validity}
For the interval in \eqref{eq:interval}, suppose the fitted quantiles and
allocation are calibration-independent.
Conditional on the data used to construct them,
the two one-sided errors satisfy
\[
\pr\{Y_{\rm new}>\widehat U_{\alpha_U}(X_{\rm new})\}\le\alpha_U,\qquad
\pr\{Y_{\rm new}<\widehat L_{\alpha_L}(X_{\rm new})\}\le\alpha_L.
\]
Consequently, if \(\alpha_L+\alpha_U=\alpha\), then
\(\pr\{Y_{\rm new}\in\widehat C_{\alpha_L,\alpha_U}(X_{\rm new})\}\ge1-\alpha\).
\end{proposition}

To see the result, condition on everything used to fit the endpoints and choose
the allocation. The \(m\) upper scores and the upper score of the test pair are
exchangeable. Its rank therefore exceeds \(k_U\) with probability at most
\(\alpha_U\); truncating the correction below at zero can only enlarge the
interval. The same argument applies to the lower scores, and a union bound gives
the total coverage statement. Thus the allocation can be estimated without using
calibration observations or altering the rank argument. The guarantee is marginal
over the new pair. Distribution-free conditional coverage given
\(X_{\rm new}=x\) generally requires additional structure
\citep{barber2021limits}.

The same rank convention includes the two one-sided endpoint cases. If
\((\alpha_L,\alpha_U)=(0,\alpha)\), omit the unused lower fitted endpoint and
report \((-\infty,\widehat U_\alpha(x)]\); its upper miss probability is at
most \(\alpha\). If \((\alpha_L,\alpha_U)=(\alpha,0)\), report
\([\widehat L_\alpha(x),+\infty)\); its lower miss probability is at most
\(\alpha\). For a nonnegative response, the first set may be intersected with
\([0,+\infty)\) to give \([0,\widehat U_\alpha(x)]\) without changing
coverage. These are the standard one-sided split-conformal sets, and their
intersection at positive tail-specific levels gives the interval in
\eqref{eq:interval} \citep{cuonzo2026tails}. The candidate allocation grid
excludes \(c=0\) and \(c=1\) because it compares finite two-sided lengths. The
rank-floor allocation is the closest finite two-sided member to the upper
one-sided set, retaining one lower calibration tail count through
\(\alpha_L=b_m\).

In the scalar case, the procedure has a more revealing representation. Let the
fitting sample have size \(n\), and define
\(\widehat\ell_s=Y^{\rm fit}_{(\lceil ns\rceil)}\) and
\(\widehat u_s=Y^{\rm fit}_{(\lceil n(1-s)\rceil)}\).
For \(j_L(s)=\lfloor(m+1)s\rfloor\) and
\(j_U(s)=\lceil(m+1)(1-s)\rceil=m+1-j_L(s)\), direct inversion of the score
order statistics gives
\begin{equation}
\label{eq:endpoint-identity}
\widehat U_s=\max\{\widehat u_s,Y^{\rm cal}_{(j_U(s))}\},
\qquad
\widehat L_s=\min\{\widehat\ell_s,Y^{\rm cal}_{(j_L(s))}\}.
\end{equation}
The right endpoint is \(+\infty\) when \(j_U(s)>m\), and the left endpoint is
\(-\infty\) when \(j_L(s)<1\). Hence each final endpoint is the more
conservative of two independent empirical quantiles, one from fitting and one
from calibration. Identity \eqref{eq:endpoint-identity} is also why the
rank floor enters the length theory: when \(s\) is of order \(1/m\),
the calibration endpoint is an extreme order statistic rather than an
intermediate one.

For scalar intervals, write \(\widehat C_{\alpha_L,\alpha_U}\) for the
interval associated with an allocation pair. When
\((\alpha_L,\alpha_U)=((1-c)\alpha,c\alpha)\), abbreviate it as
\(\widehat C_c\). Also write \(\widehat C_{\rm ET}=\widehat C_{1/2}\) for the
equal-tail interval and
\(\widehat C_{\rm RF}=\widehat C_{1-b_m/\alpha}\) for the rank-floor interval.
Thus allocation appears in the subscript, while a covariate argument, when
present, appears in parentheses.

The exact max-min representation links the classical population
shortest-interval problem to the conformally calibrated interval. It reduces
calibrated length to the joint behavior of independent fitting and calibration
order statistics at the two allocated tail probabilities. The tail-scale
relations and finite-rank limits below are consequences of this representation.

\subsection{The empirical allocation criterion}
\label{sec:selection}

Let \(q_p(x)\) denote the conditional quantile of \(Y\) given \(X=x\). The
population width associated with a global allocation fraction \(c\) is
\begin{equation}
\label{eq:population-width}
\mathcal J_\alpha(c)=
\E\{q_{1-c\alpha}(X)-q_{(1-c)\alpha}(X)\}.
\end{equation}
In the scalar setting this reduces to
\(q_{1-c\alpha}-q_{(1-c)\alpha}\), the usual lower-tail parameterization of
a connected probability interval. Minimizing it is the classical shortest
interval problem \citep{liu2015spin,brehmer2021intervals}. The present
selection criterion is a sample version of \eqref{eq:population-width}; the
new issue is whether its allocation continues to control length after both
endpoints have been estimated and conformally calibrated.

The criterion selects one global allocation fraction by averaging fitted width over the
selection covariates. This is appropriate when the direction and relative
strength of tail asymmetry are stable over the covariate space, and it retains
one explicit tail-specific miscoverage level for each side. Section
\ref{sec:heterogeneity} quantifies the efficiency loss when covariate groups
have opposite dominant tails.

The candidate allocation grid has both a statistical and a finite-sample role.
The grid \(\mathcal G_m\) defined in Section \ref{sec:construction} spans the
full finite-rank allocation domain and contains both allocation-domain endpoints. In the
dominant-upper-tail branch, the empirical allocation rule can approach the
right endpoint. Comparable tails make the fitted width diverge near both
endpoints and keep the minimizer in the interior. The same grid therefore
adapts to both regularly varying branches without a preliminary
classification. Candidates with
\(c\alpha<b_m\) or \((1-c)\alpha<b_m\) give a zero calibration tail count
and produce an interval with an infinite endpoint.

Separate calibration assigns a distinct guarantee to each error direction. The upper
miss probability is controlled by \(\alpha_U\) and the lower miss probability
by \(\alpha_L\), regardless of which candidate is selected on the selection sample.
Coverage and length are then separate questions: Proposition
\ref{prop:validity} gives the former, and Sections \ref{sec:theory} and
\ref{sec:covariates} characterize the latter.

\section{Tail-quantile response and calibration resolution}
\label{sec:theory}

Changing the upper-tail allocation fraction replaces an endpoint quantile at
tail probability \(\alpha\) by one at \(c\alpha\). We first classify the
resulting length change through the response of the tail quantile to this
proportional rescaling. Regular variation yields a multiplicative response,
while de Haan variation yields an additive response on a smaller auxiliary
scale. We then vary \(m\alpha\), which determines whether ordinary conformal
ranks are asymptotically smooth or remain visible in the first-order limit.
Tail-quantile response and calibration resolution are distinct.

\subsection{Regularly varying tails}
\label{sec:frechet}

We first suppress covariates and use empirical quantile endpoints. The
miscoverage level \(\alpha=\alpha_{n,m}\) now tends to zero as the fitting and
calibration sample sizes increase. For the asymptotic results, the fitting and
calibration observations are two independent random samples from \(F\). Let \(F\) be
continuous, let \(q_p=F^{-1}(p)\), and define the upper and lower tail scales
by \(U(s)=q_{1-s}\) and \(D(s)=(-q_s)_+\).
Suppose \(U\) is regularly varying at zero with index \(-\xi\), where
\(\xi>0\): for every fixed \(t>0\),
\(U(ts)/U(s)\to t^{-\xi}\). The exponent \(\xi\) measures how strongly the
upper endpoint reacts to a proportional change in tail probability.

Two lower-tail branches lead to different allocations. In the
dominant-upper-tail branch, \(|q_s|\le A_Ls^{-\xi_L}\) for all sufficiently
small \(s\), for some \(A_L<\infty\) and \(0\le\xi_L<\xi\), and
\(b_m^{-\xi_L}=o\{U(\alpha)\}\).
The lower endpoint at the rank floor is then negligible on the upper
scale. In the comparable-tail branch,
\(D(s)/U(s)\to\gamma\in(0,\infty)\), so both endpoints have the same tail
index and \(\gamma\) is the lower-to-upper tail-scale ratio. The two branches
meet continuously at \(\gamma=0\), but their
optimal allocations differ: the first has a boundary optimum and the second an
interior optimum.

The first-order results use the intermediate-rank conditions
\(n\alpha\to\infty\), \(m\alpha\to\infty\), and \(b_m=o(\alpha)\).
The first two conditions make the working upper order statistics intermediate
rather than extreme. The last leaves room between the rank floor and
the target miscoverage level. Together they define the intermediate-rank
regime.
For example, if \(n\) and \(m\) have the same order,
\(U(\alpha)\) has order \(\alpha^{-\xi}\), \(b_m\) has order \(m^{-1}\),
and \(\alpha=m^{-\chi}\), the dominant-upper-tail conditions reduce to
\(\xi_L/\xi<\chi<1\).
Thus the analysis concerns tail probabilities that shrink with sample size
while leaving an increasing number of calibration observations in each target
tail. Regular-variation facts used below
follow from \citet{bingham1987regular}; the rank arguments are proved in the
Supplementary Material.

\begin{theorem}
\label{thm:length-relation}
For deterministic allocations with \(\alpha_U/\alpha\to c\in(0,1)\),
\[
\frac{\widehat U_{\alpha_U}}{U(\alpha)}\xrightarrow{\rm p}c^{-\xi},
\qquad
\frac{-\widehat L_{\alpha_L}}{U(\alpha)}\xrightarrow{\rm p}
\gamma(1-c)^{-\xi},
\]
and hence
\[
\frac{|\widehat C_{\alpha_L,\alpha_U}|}{U(\alpha)}
\xrightarrow{\rm p}
g_\gamma(c)=c^{-\xi}+\gamma(1-c)^{-\xi},
\]
where \(\gamma=0\) in the dominant-upper-tail branch. If \(\gamma=0\), the
same conclusion extends to the moving boundary
\((\alpha_L,\alpha_U)=(b_m,\alpha-b_m)\), with limit one. If
\(\gamma>0\), the normalized length diverges when \(c\) tends to either
endpoint of the unit interval.
\end{theorem}

Identity \eqref{eq:endpoint-identity} explains the two terms in the limit. The
upper calibrated endpoint is a maximum of independent fitting and calibration
order statistics at level \(\alpha_U\); the lower endpoint is the analogous
minimum at level \(\alpha_L\). Under \(m\alpha\to\infty\), both order statistics are
consistent for the same population quantile. Regular variation then converts
\(U(c\alpha)\) to \(c^{-\xi}U(\alpha)\), and the lower scale contributes
\(\gamma(1-c)^{-\xi}U(\alpha)\). The main proof difficulty is the moving
boundary \(c\to1\): the lower rank approaches the rank floor,
so pointwise intermediate-order consistency is unavailable. A uniform
order-statistic bound shows that this calibrated lower endpoint remains
\(o_{\rm p}\{U(\alpha)\}\).

\begin{figure}[!htb]
\centering
\includegraphics[width=0.70\textwidth]{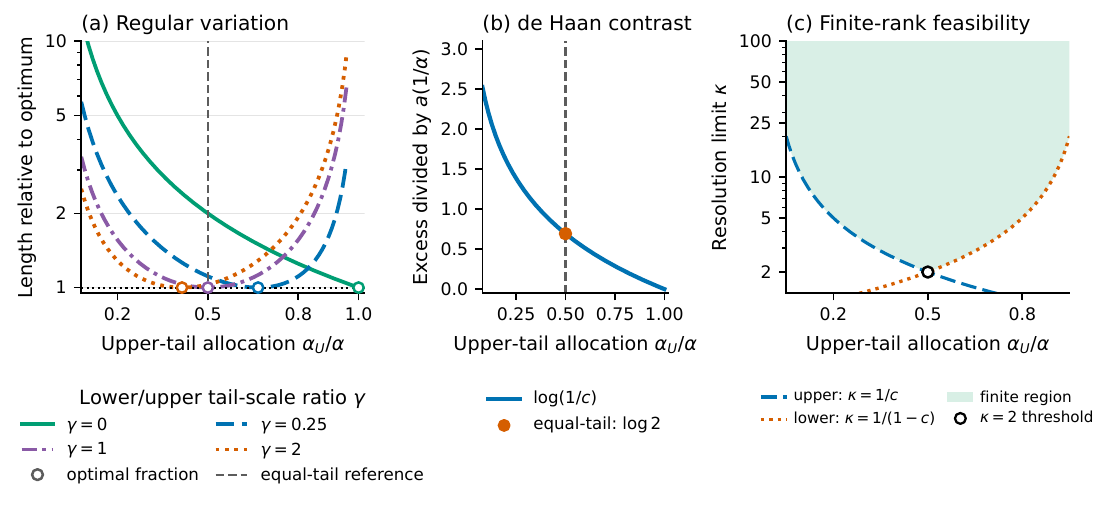}
\caption{Three forms of tail-allocation geometry at \(\xi=1\).}
\alttext{Three panels show the regularly varying allocation-length function and its optimum, the de Haan additive allocation contrast, and the two-sided finite-rank feasibility region.}
\label{fig:geometry}
\end{figure}

The allocation-length function \(g_\gamma\) has a unique optimum. For
\(\gamma>0\),
\begin{equation}
\label{eq:cstar}
c^\ast=\{1+\gamma^{1/(\xi+1)}\}^{-1},\qquad
g_\gamma(c^\ast)=\{1+\gamma^{1/(\xi+1)}\}^{\xi+1}.
\end{equation}
The equal-tail constant is \(g_\gamma(1/2)=2^\xi(1+\gamma)\), so its ratio to
the optimum is
\begin{equation}
\label{eq:equal-ratio}
\frac{g_\gamma(1/2)}{g_\gamma(c^\ast)}
=\frac{2^\xi(1+\gamma)}
{\{1+\gamma^{1/(\xi+1)}\}^{\xi+1}}\ge1.
\end{equation}
The ratio in \eqref{eq:equal-ratio} equals one exactly when \(\gamma=1\). As
\(\gamma\downarrow0\), \(c^\ast\to1\) and the ratio tends to \(2^\xi\).
Equal tails are therefore first-order optimal when the tail scales match and
become increasingly inefficient as one side becomes dominant. When both tails
remain heavy, the normalized length
instead diverges at the rank-floor allocation. Figure \ref{fig:geometry}
previews these three geometries.

\FloatBarrier

Thus every deterministic sequence whose upper-tail allocation fraction converges to \(c\) has
the corresponding first-order length constant \(g_\gamma(c)\); starving either
comparable-tail branch makes the normalized length diverge. Consequently, an
allocation attains the smallest first-order constant only if its upper-tail
allocation fraction converges to the unique minimizer in \eqref{eq:cstar}. The next result
extends that sharpness statement to calibration-independent random allocations and
connects it to the empirical allocation rule in Section \ref{sec:selection}.

\subsection{Empirical allocation rule and sharpness}
\label{sec:adaptive-theory}

Use the candidate allocation grid \(\mathcal G_m\) defined in Section
\ref{sec:construction}. For scalar empirical quantiles, the empirical
allocation rule uses
\(T_n(c)=\widehat u_{c\alpha}-\widehat\ell_{(1-c)\alpha}\) and chooses
\(\widehat c\in\operatorname*{argmin}_{c\in\mathcal G_m}T_n(c)\).
The minimum over the finite grid is attained; ties are broken by taking the
smallest minimizer.

\begin{theorem}
\label{thm:adaptive}
Let \(\widetilde c\in\mathcal C_m(\alpha)\) be any
calibration-independent allocation fraction.
The resulting interval retains the coverage guarantee of Proposition
\ref{prop:validity}. For every
\(\varepsilon>0\), no such rule can improve on the first-order optimum by a fixed amount:
\[
\pr\left\{
\frac{|\widehat C_{\widetilde c}|}{U(\alpha)}
<g_\gamma(c^\ast)-\varepsilon\right\}\longrightarrow0,
\]
where \(c^\ast=1\) and \(g_0(c^\ast)=1\) in the dominant-upper-tail branch. Moreover,
\[
\frac{|\widehat C_{\widetilde c}|}{U(\alpha)}
\xrightarrow{\rm p}g_\gamma(c^\ast)
\quad\text{if and only if}\quad
\widetilde c\xrightarrow{\rm p}c^\ast.
\]
For the empirical allocation rule \(\widehat c\) over \(\mathcal G_m\),
this condition holds, and therefore
\[
\frac{|\widehat C_{\widehat c}|}{U(\alpha)}
\xrightarrow{\rm p}
\begin{cases}
1, & \gamma=0,\\
g_\gamma(c^\ast), & \gamma>0.
\end{cases}
\]
In the second case, \(\widehat c\xrightarrow{\rm p}c^\ast\).
\end{theorem}

Monotonicity of the endpoint order statistics and continuity of
\(g_\gamma\) upgrade the pointwise relation in Theorem \ref{thm:length-relation} to
uniform convergence on fixed interior intervals. When \(\gamma>0\), divergence
of the upper-endpoint term rules out a strip at the left edge, and
divergence of the lower-endpoint term rules out a strip at the right edge.
The empirical allocation rule therefore remains in an interior interval
containing \(c^\ast\), where strict convexity identifies it. When \(\gamma=0\),
the upper endpoint excludes the left edge and the full grid reaches the
boundary optimum at \(c^\ast=1\). The random-allocation argument also shows that attaining
the optimal length constant is equivalent to selecting the optimal allocation fraction.

The empirical allocation rule therefore requires no estimate of \(\xi\) or
\(\gamma\), and no preliminary choice between the two regularly varying
branches. Its selection consistency covers both branches, while Section
\ref{sec:gumbel} identifies a smaller additive allocation contrast in the de
Haan class.

\subsection{Optimality beyond allocated intervals}
\label{sec:oracle-optimality}

The allocation optimum can also be compared with prediction sets outside the
one-sided interval family. This requires expected length rather than a
pathwise comparison, because marginal validity permits randomized sets that are
occasionally empty and occasionally unbounded.

\begin{proposition}
\label{prop:oracle}
Suppose the dominant-upper-tail conditions hold and \(Y\) has a density \(f\) that is
positive and nonincreasing beyond a finite threshold. Define
\(t_\alpha=\inf\{t>0:\pr\{f(Y)>t\}\le1-\alpha\}\), and suppose the level sets
\(\{y:f(y)=t_\alpha\}\) have zero Lebesgue measure for all sufficiently small
\(\alpha\). Let \(C\) be any random measurable set based
on data independent of a new scalar response, with
\(\pr(Y_{\rm new}\in C)\ge1-\alpha\). Then
\[
\E\{\operatorname{Leb}(C)\}\ge U(\alpha)\{1-o(1)\}.
\]
The rank-floor interval and the interval from the empirical allocation rule satisfy
\(|\widehat C|/U(\alpha)\to1\) in probability. If their normalized lengths are
uniformly integrable, then
\[
\E(|\widehat C|)=U(\alpha)\{1+o(1)\},
\]
so they attain the lower bound to first order.
\end{proposition}

The lower bound is the density-level-set argument used in optimal
prediction-set theory \citep{lei2014prediction}. Among all inclusion functions
with at least \(1-\alpha\) probability mass, a highest-density region minimizes
Lebesgue measure. Monotonicity of the upper density forces such a region to
extend at least to \(U(\alpha)\), up to a fixed lower endpoint, which gives the
displayed scale. Uniform integrability is an additional condition:
convergence in probability of normalized length alone does not control its
expectation under very heavy tails. The allocation optimality results in
Sections \ref{sec:frechet} and \ref{sec:adaptive-theory} require no density
assumption; expected-length optimality among arbitrary valid sets additionally
requires the density and uniform-integrability conditions above.

\subsection{The de Haan auxiliary-scale boundary}
\label{sec:gumbel}

Regularly varying endpoints respond multiplicatively to a fixed proportional
change in tail probability. This mechanism disappears when
\(U(c\alpha)/U(\alpha)\to1\), but allocation can remain visible on a smaller
additive scale.

Let \(V(t)=U(1/t)\). We write \(V\in\Pi(a)\) when
\(\{V(tx)-V(t)\}/a(t)\to\log x\) for every \(x>0\), where \(a(t)>0\) is an
auxiliary function. This class is a standard representation of Gumbel-domain upper quantiles
\citep{dehaan2006extreme}. Its auxiliary function is slowly varying and
\(a(t)/V(t)\to0\). Thus \(V\) is slowly varying, but the auxiliary function
\(a\) preserves the first nonvanishing effect of changing the allocation
fraction.

\begin{theorem}
\label{thm:pi}
Suppose \(V\in\Pi(a)\) is unbounded, \(n\alpha\to\infty\),
\(m\alpha\to\infty\), and
\(\alpha_U/\alpha\to c\in(0,1]\). If the calibrated lower endpoint is
\(o_{\rm p}\{a(1/\alpha)\}\), then
\[
\frac{|\widehat C_{\alpha_L,\alpha_U}|-U(\alpha)}
{a(1/\alpha)}
\xrightarrow{\rm p}\log(1/c).
\]
For two fixed positive allocation fractions \(c_a\) and \(c_b\) satisfying the same
lower-endpoint condition,
\[
\frac{|\widehat C_{c_a}|-|\widehat C_{c_b}|}{a(1/\alpha)}
\xrightarrow{\rm p}\log(c_b/c_a).
\]
In particular, the equal-tail interval exceeds the rank-floor interval by
\(a(1/\alpha)\{\log 2+\op(1)\}\), while their length ratio converges to one.
\end{theorem}

The lower-endpoint condition is on the \(a\)-scale and is stronger than
negligibility relative to \(U(\alpha)\). It isolates a one-sided upper-tail
comparison. A symmetric Gaussian or Laplace distribution,
for example, generally requires both additive endpoint contributions. For a
positive lognormal response the condition holds automatically: the calibrated
lower endpoint is bounded in probability, while \(a(1/\alpha)\to\infty\).

The auxiliary function organizes several familiar examples. Its size
determines whether the equal-tail excess diverges, converges to a positive
constant, or vanishes.
For a lognormal response with log-scale standard deviation \(\sigma\), let
\(z(t)=\Phi^{-1}(1-1/t)\). Then
\[
\frac{a(1/\alpha)}{U(\alpha)}\sim\frac{\sigma}{z(1/\alpha)},
\qquad
\frac{|\widehat C_{\rm ET}|}{|\widehat C_{\rm RF}|}
=1+\frac{\sigma\log2}{\sqrt{2\log(1/\alpha)}}\{1+\op(1)\}.
\]
The relative gain therefore vanishes slowly even when the finite-sample
difference is appreciable. For a one-sided Weibull-type upper tail
\(1-F(y)=\exp(-y^\beta)\),
\[
U(\alpha)=\{\log(1/\alpha)\}^{1/\beta},
\qquad
a(1/\alpha)=\frac{1}{\beta}
\{\log(1/\alpha)\}^{1/\beta-1}.
\]
The absolute equal-tail excess diverges for \(\beta<1\), equals \(\log2\)
for the exponential case \(\beta=1\), and vanishes for \(\beta>1\). For a
Gaussian-type upper tail with a negligible lower endpoint, even the absolute
excess vanishes. These examples show why allocation displacement and
first-order relative improvement are distinct phenomena.

\subsection{Finite-rank regime}
\label{sec:resolution}

The preceding results require \(m\alpha\to\infty\), so calibration noise is
lower order on the tail-quantile scale. In the finite-rank regime,
\(m\alpha\to\kappa\in(0,\infty)\), only finitely many calibration observations
fall in the target tails. Rank discreteness then enters the first-order length.
Fitting-sample quantiles can still be consistent if \(n\alpha\to\infty\); these two
conditions imply \(n/m\to\infty\), separating endpoint estimation from
finite-rank calibration.

Define the available upper and lower calibration tail counts by
\(h_U=\lfloor(m+1)\alpha_U\rfloor\) and
\(h_L=\lfloor(m+1)\alpha_L\rfloor\).
Identity \eqref{eq:endpoint-identity} shows that the right endpoint is finite
exactly when \(h_U\ge1\), and the left endpoint is finite exactly when
\(h_L\ge1\). Thus a two-sided interval requires two positive calibration tail counts, even when
one tail is negligible for first-order length when \(m\alpha\to\infty\).

\begin{theorem}
\label{thm:resolution}
Suppose \(m\alpha\to\kappa\in(0,\infty)\),
\(n\alpha\to\infty\), and \(\alpha_U/\alpha\to c\in(0,1)\). If either
\(h_U=0\) or \(h_L=0\) eventually, the interval has infinite length. If the
calibration tail counts converge to positive integers \(h_U^\star\) and \(h_L^\star\), then the following
limits hold.

In the dominant-upper-tail branch,
\[
\frac{|\widehat C|}{U(\alpha)}\xrightarrow{\rm d}
\max\left\{c^{-\xi},
\left(\frac{\kappa}{\Gamma_{h_U^\star}}\right)^\xi\right\}.
\]
In the comparable-tail branch,
\[
\frac{|\widehat C|}{U(\alpha)}\xrightarrow{\rm d}
\max\left\{c^{-\xi},
\left(\frac{\kappa}{\Gamma_{h_U^\star}}\right)^\xi\right\}
+\gamma\max\left\{(1-c)^{-\xi},
\left(\frac{\kappa}{\Gamma'_{h_L^\star}}\right)^\xi\right\},
\]
where \(\Gamma_h\) and \(\Gamma'_h\) are independent gamma random variables
with shape \(h\) and unit rate. If
\(\kappa c\notin\mathbb Z\) and \(\kappa(1-c)\notin\mathbb Z\), then
\(h_U\to\lfloor\kappa c\rfloor\) and
\(h_L\to\lfloor\kappa(1-c)\rfloor\). Under these conditions, the interval is
eventually finite exactly when \(c\in(1/\kappa,1-1/\kappa)\), and this
finite-rank allocation domain is nonempty exactly when \(\kappa>2\).
\end{theorem}

The gamma variables are the arrival times of the limiting upper and lower
Poisson tail processes. Each maximum has a direct interpretation. The first
entry is the fitted endpoint; the second is the finite-rank calibration
endpoint. Calibration therefore remains visible even when fitting-sample quantiles
are consistent. In the comparable-tail branch the two tail processes are
asymptotically independent because upper and lower exceedances occupy disjoint
rare-event cells. In the dominant-upper-tail branch the lower endpoint is still required
for finiteness, but its magnitude is negligible on the upper scale.

The feasibility threshold \(\kappa=2\) is distinct from the smooth
allocation-length
function in Theorem \ref{thm:length-relation}. Below it, no fixed allocation fraction leaves at least
one positive calibration tail count on each side. Above it, the finite-rank allocation domain expands with
\(\kappa\). Letting \(\kappa\to\infty\) in the limit law, with
\(h_U/\kappa\to c\) and \(h_L/\kappa\to1-c\), the gamma ratios concentrate
and the random limit contracts to \(g_\gamma(c)\), recovering the
intermediate-rank result. Coverage remains the
finite-sample guarantee in Proposition \ref{prop:validity}; infinite endpoints
arise from the ordinary rank convention rather than from an extrapolation
model.

Table \ref{tab:classification} separates the effect of tail-quantile response
from the effect of calibration resolution. In the regularly varying rows, \(\gamma\) is the
lower-to-upper tail-scale ratio. The de Haan row assumes diverging calibration
tail counts and a lower endpoint that is negligible on the auxiliary scale.
The finite-rank row concerns the regularly varying branches and presumes
positive calibration tail counts on both sides. Rank feasibility also remains
relevant for de Haan tails, while its bounded-count fluctuation would occur on
the auxiliary scale and is not developed here.

\FloatBarrier
\begin{table}[!ht]
\caption{Allocation effects by tail-quantile response and calibration resolution.}
\label{tab:classification}
\centering
\begin{tabular}{>{\raggedright\arraybackslash}p{0.27\textwidth}
>{\raggedright\arraybackslash}p{0.30\textwidth}
>{\raggedright\arraybackslash}p{0.31\textwidth}}
\toprule
Setting & Natural limit & Allocation consequence \\
\midrule
\multicolumn{3}{l}{\textit{Tail-quantile response}} \\
\addlinespace
Dominant Fr\'echet tail
& \(|\widehat C_c|/U(\alpha)\to c^{-\xi}\)
& Boundary optimum \(c^\ast=1\); the equal-tail length ratio is \(2^\xi\). \\
Comparable Fr\'echet tails
& \(|\widehat C_c|/U(\alpha)\to
c^{-\xi}+\gamma(1-c)^{-\xi}\)
& Interior optimum in \eqref{eq:cstar}; equal tails are optimal exactly when
\(\gamma=1\). \\
de Haan \(\Pi\) class with negligible lower endpoint
& \(\{|\widehat C_c|-U(\alpha)\}/a(1/\alpha)\to\log(1/c)\)
& Allocation has an additive effect, while every fixed positive allocation fraction has
the same first-order relative scale. \\
\addlinespace
\multicolumn{3}{l}{\textit{Calibration resolution}} \\
\addlinespace
Bounded tail counts, \(m\alpha\to\kappa\)
& Sum of fitted-endpoint constants and Poisson order-statistic maxima
& Both calibration tail counts constrain finiteness; the window for fixed allocation fractions opens only
when \(\kappa>2\). \\
\bottomrule
\end{tabular}
\end{table}

Two refinements are given in the Supplementary Material. A functional
fluctuation result describes the maximum of the fitting and calibration tail
quantile processes over \(c\); the fitting-to-calibration sample-size ratio
enters this limit and explains part of the finite-sample inflation above the
first-order constants. A second-order calculation shows that the exact
population optimum in the dominant-upper-tail branch is interior, although its
first-order fraction tends to one.

\FloatBarrier

\section{Covariates and tail heterogeneity}
\label{sec:covariates}

The scalar geometry transfers over covariates when the tail direction and
scale structure are homogeneous. When the dominant direction changes across
groups, no common allocation fraction can retain groupwise first-order
efficiency.

\subsection{Transfer under tail homogeneity}
\label{sec:transfer}

Proposition \ref{prop:validity} already allows arbitrary covariates and fitted
quantile functions. Transferring the length result requires additional
structure. Suppose
\(Y=\mu(X)+\sigma(X)\epsilon\), where
\(0<\sigma_{\min}\le\sigma(X)\le\sigma_{\max}<\infty\) and
\(\epsilon\) is independent of \(X\). Its upper quantile
\(U_\epsilon(s)\) is regularly varying with index \(-\xi\), and its lower
tail satisfies \(|q_{\epsilon,s}|\le A_Ls^{-\xi_L}\) for some
\(0\le\xi_L<\xi\). Then
\(q_{1-\alpha_U}(x)=\mu(x)+\sigma(x)U_\epsilon(\alpha_U)\) and
\(q_{\alpha_L}(x)=\mu(x)+\sigma(x)q_{\epsilon,\alpha_L}\). Assume
\[
\sup_x\frac{|\widehat q_{1-\alpha_U}(x)-q_{1-\alpha_U}(x)|}
{\sigma(x)U_\epsilon(\alpha_U)}\xrightarrow{\rm p}0,
\qquad
\sup_x|\widehat q_{\alpha_L}(x)-q_{\alpha_L}(x)|
=o_{\rm p}\{U_\epsilon(\alpha)\}.
\]
These conditions encode the endpoint accuracy supplied by the extreme
conditional quantile estimator; their verification depends on the estimator
and the dimension of \(X\). Breiman's product-tail result gives a related marginal
extension when the scale is unbounded but satisfies a suitable moment condition
\citep{breiman1965limit}.
Assume also that \(n\alpha\to\infty\), \(m\alpha\to\infty\),
\(\alpha_L\ge b_m=1/(m+1)\), \(b_m=o(\alpha)\), and
\(b_m^{-\xi_L}=o\{U_\epsilon(\alpha)\}\), with
\(\alpha_L+\alpha_U=\alpha\). These conditions make both one-sided
calibration corrections negligible on the upper-tail scale when the upper-tail
miscoverage level remains a positive fraction of \(\alpha\).

\begin{theorem}
\label{thm:transfer}
Under the tail-homogeneous model, the uniform endpoint consistency conditions,
and the regime conditions above, deterministic allocations with
\(\alpha_U/\alpha\to c\in(0,1]\) satisfy
\[
\sup_x\left|
\frac{\len\{\widehat C_{\alpha_L,\alpha_U}(x)\}}
{\sigma(x)U_\epsilon(\alpha)}-c^{-\xi}
\right|\xrightarrow{\rm p}0.
\]
The interval retains the marginal coverage statement in Proposition
\ref{prop:validity}.
\end{theorem}

The common upper-tail allocation fraction appears as a multiplicative constant at every
covariate value. Hence the equal-tail length ratio \(2^\xi\) transfers uniformly, and
the average conditional length has leading term
\(c^{-\xi}U_\epsilon(\alpha)\E\{\sigma(X)\}\). If, in addition, the
selection criterion \(T_{n_{\rm se}}(c)\) converges uniformly to this average core length
over candidate allocation grids for which the limiting allocation-length function has a unique optimum,
the ordinary argmin argument
gives \(\widehat c\to c^\ast\) and the selected interval attains the corresponding
constant. A selected-allocation result therefore additionally requires uniform
convergence of the normalized selection criterion.

\subsection{Opposite dominant tails across groups}
\label{sec:heterogeneity}

Tail homogeneity concerns direction as well as index. Consider two equally
likely groups. In the first, the upper tail has index \(\xi\) and the lower
tail is lighter. In the second, the lower tail has index \(\xi\) and the upper
tail is lighter. Let
\(\len_j^\ast(\alpha)\) be the shortest population interval length in group
\(j\). For a deterministic global allocation pair
\(\alpha_L+\alpha_U=\alpha\), let \(\len_j\) denote the length in group
\(j\) obtained from the oracle group quantiles and arbitrary nonnegative
endpoint corrections. Write \(c=\alpha_U/\alpha\), and interpret
\(0^{-\xi}=+\infty\).

\begin{proposition}
\label{prop:opposite-tails}
For any such global allocation sequence satisfying \(c\to c_0\in[0,1]\), the
following bound holds for every realization of the nonnegative corrections:
\[
\liminf_{\alpha\downarrow0}\max_{j=1,2}
\frac{\len_j}{\len_j^\ast(\alpha)}
\ge \max\{c_0^{-\xi},(1-c_0)^{-\xi}\}\ge2^\xi.
\]
Population equal-tail intervals attain the group-minimax constant \(2^\xi\),
whereas group-dependent oracle allocations attain one in both groups.
\end{proposition}

Proposition \ref{prop:opposite-tails} identifies the role of directional tail
homogeneity in Theorem \ref{thm:transfer}. A common allocation can be efficient
across covariate values only when the dominant direction is stable; opposite
directions require group-specific allocation and calibration.

\section{Numerical evidence}
\label{sec:numerical}

\subsection{Tail classes and the empirical allocation rule}
\label{sec:simulation-scalar}

The simulations follow the theory in the same order. The scalar study examines
the regularly varying constants, the empirical allocation rule, and the de Haan
additive contrast; the covariate study then examines transfer and groups with
opposite dominant tails.
The scalar study uses six continuous distributions: two piecewise-quantile
dominant-upper-tail models, a comparable-tail model with
\(\gamma=2\), Student \(t_3\), lognormal with unit log-scale standard
deviation, and standard normal. In the piecewise models, the lower and upper
tail indices are \((0.2,1.0)\), \((0.5,1.5)\), and \((1,1)\), respectively;
the third model has lower-to-upper tail-scale ratio two. The central 60 percent of
each piecewise quantile function is linearly interpolated between its tail
pieces. We use miscoverage levels from 0.05 to 0.002, equal fitting and calibration
sizes of \(2{,}000\), \(10{,}000\), and \(50{,}000\), and 500 replications.

The main comparison uses the equal-tail allocation, rank-floor allocation,
empirical allocation rule, and population-optimal allocation, all under separate one-sided calibration and
with common fitting and calibration samples. The Supplementary Material gives
full definitions, an uncalibrated reference, and joint numerical summaries.
The empirical allocation rule uses a candidate allocation grid with spacing
0.005.

\begin{figure}[!htb]
\centering
\includegraphics[width=0.86\textwidth]{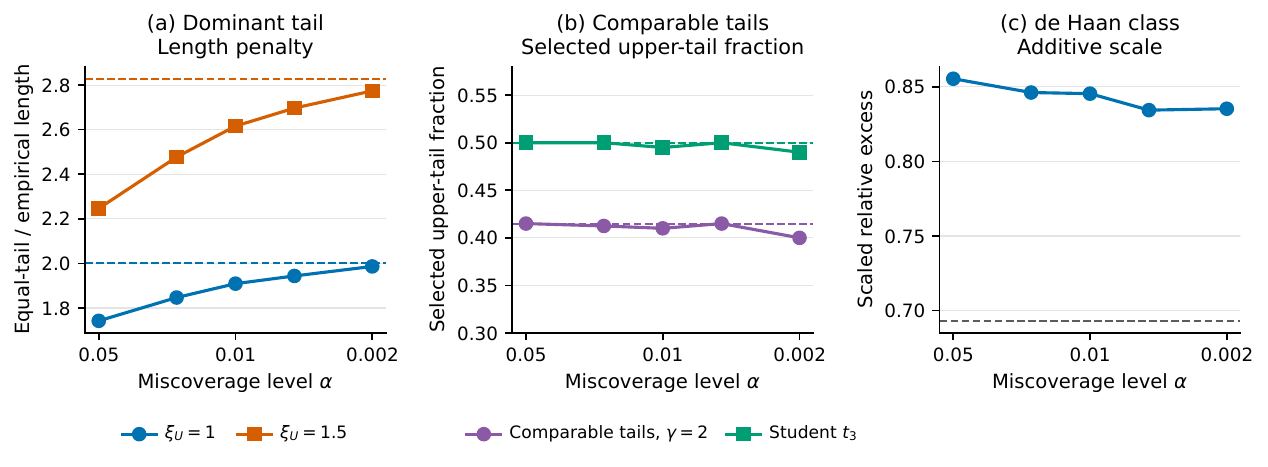}
\caption{Tail-class simulations at \(n=m=50{,}000\); dashed lines show
theoretical limits.}
\alttext{Three panels compare simulated tail-allocation behavior with theory. In the first, two ratios of equal-tail length to the length from the empirical allocation rule rise toward their heavy-tail limits as miscoverage decreases. In the second, selected upper-tail allocation fractions remain near the theoretical optima for comparable tails and Student t data. In the third, the scaled lognormal excess approaches a horizontal logarithmic reference slowly.}
\label{fig:tail-regimes}
\end{figure}
\FloatBarrier

In Fig. \ref{fig:tail-regimes}(a), the ratio of equal-tail length to the length
from the empirical allocation rule in the
dominant-upper-tail models approaches 2 and
\(2^{1.5}\), respectively. In the comparable-tail model, the median selected
allocation fraction remains near \(1/(1+\sqrt 2)=0.414\); for Student \(t_3\), it remains
near one half, as shown in panel (b). The lognormal scaled excess in panel (c)
stays above \(\log 2\) at the
displayed sample sizes, which is consistent with the slow remainder quantified
by the lognormal calculation following Theorem \ref{thm:pi}, rather than with a
Fr\'echet constant.
The same study shows the cost of forcing the rank-floor allocation when both
tails remain heavy. For the \(\gamma=2\) model at \(\alpha=0.01\) and
\(n=m=50{,}000\), its median normalized length is \(2.84\times10^3\), compared
with 6.23 for equal-tail and 6.02 for the empirical allocation rule.

\subsection{Covariate transfer and opposite dominant tails}
\label{sec:simulation-covariate}

The covariate study uses \(X\) uniform on the unit interval,
\(\mu(x)=2x-1\), \(\sigma(x)=1+x\), and the first dominant-upper-tail noise
distribution above. We compare oracle conditional quantiles with estimated
location-scale quantiles based on median and interquartile regressions and
standardized residual order statistics. The sample sizes are
\(n=m=20{,}000\), the miscoverage levels are 0.02 and 0.01, and 300 replications are
used. A separate two-group experiment assigns the dominant upper tail to one
group and the dominant lower tail to the other at miscoverage level 0.01.

\begin{figure}[!htb]
\centering
\includegraphics[width=\textwidth]{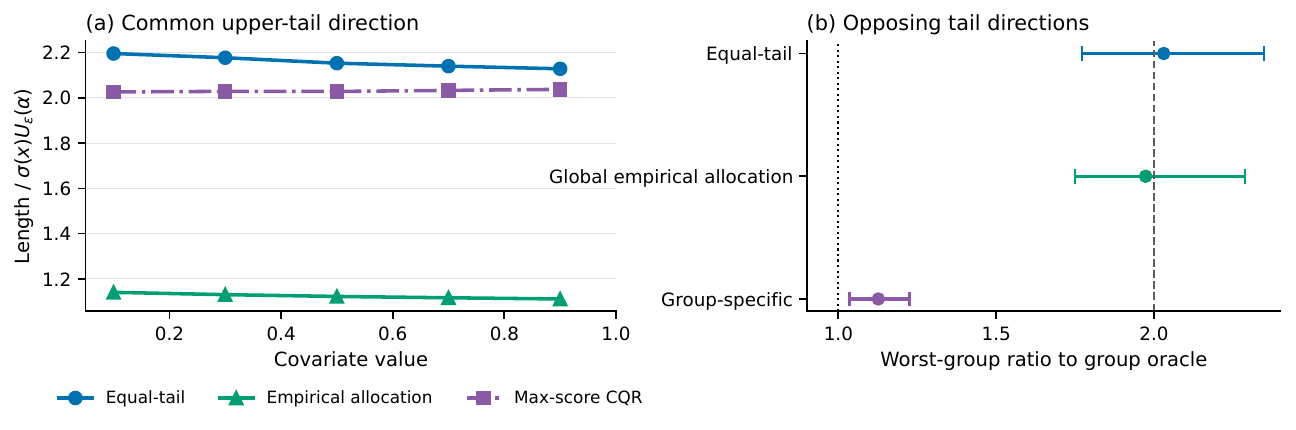}
\caption{Covariate simulations under (a) tail homogeneity and (b) opposite
dominant tails. Curves and points are medians; bars span the 10th to 90th
percentiles.}
\alttext{The first panel shows nearly flat normalized length curves over the covariate, with the empirical allocation rule close to one and equal-tail near two. The second shows median worst-group length ratios and percentile bars: equal-tail and the global empirical allocation rule are near the minimax value two, and group-specific allocation is near one.}
\label{fig:covariates}
\end{figure}
\FloatBarrier

The transfer relation is visible in Fig. \ref{fig:covariates}(a): normalized
length is nearly constant in \(x\), and estimated intervals from the empirical allocation rule remain
close to the dominant-upper-tail optimum. Across the displayed methods and fitting
modes at \(\alpha=0.01\), marginal coverage rounds to values between 0.9900
and 0.9905.
A five-covariate location-scale sensitivity with estimated endpoints gives the
same allocation and length pattern; its full design and numerical summary are
reported in the Supplementary Material.
Panel (b) matches the heterogeneity bound. Equal-tail and the global empirical
allocation rule have maximum group ratios near 2. Rank-floor allocation is
omitted from the panel because its worst-group ratio is far outside the
displayed scale: its median is 279, with 10th and 90th percentiles 85 and
1385. Group-specific allocation approaches one but uses group-specific
calibration and therefore represents a different inferential construction.
The dotted and dashed lines mark the group oracle at one and the global lower
bound at two.

\section{Applications}
\label{sec:applications}

Two nonnegative, right-skewed responses illustrate how directional misses and
finiteness complement coverage and length. Neither application prescribes
tail-specific levels, so the division of miscoverage remains available for
length. The empirical allocation rule then selects the rank-floor allocation,
the upper endpoint of the finite-rank allocation domain and the closest finite
two-sided counterpart of the one-sided interval
\([0,\widehat U_\alpha(x)]\).

\subsection{Data, targets, and comparison}
\label{sec:application-design}

The first application predicts total claim amount among claim-positive
policies in the French motor third-party liability data distributed with
\texttt{CASdatasets} \citep{dutang2026casdatasets}. Repeated claims are
aggregated to 24{,}944 policy totals. The predictors describe exposure,
vehicle, driver, and region; the contemporaneous claim count is excluded.
The response is nonnegative and has a compressed lower tail together with a
much more dispersed upper tail.

The second application predicts final wildfire size in acres using the 2020
FPA FOD-Attributes data \citep{pourmohamad2024wildfire,pourmohamad2023data}.
Restriction to the contiguous United States and removal of records with sentinel codes
leave 72{,}685 events. Predictors available at or near discovery describe
timing and location, ownership, topography, land cover, population, human
modification, and climate normals. Final-size classes, containment information,
and post-ignition summaries are excluded. The median size is 0.35 acres, while
the 99.9th percentile is 22{,}957 acres. Both applications therefore
combine a compressed lower endpoint with a much more dispersed upper endpoint.

Each of 20 paired splits assigns 40 percent of observations to quantile fitting,
10 percent to allocation selection, 25 percent to calibration, and 25 percent to
evaluation. A quantile regression forest is fitted to the logarithm
of the response, and its fitted quantiles are transformed back before calibration
\citep{meinshausen2006quantile}. The comparison separates allocation from score
construction. The equal-tail allocation and empirical allocation rule use the same one-sided
calibration, max-score conformalized quantile regression uses a common score
\citep{romano2019conformalized}, and exact distributional conformal prediction
represents a distributional shortest-interval construction
\citep{chernozhukov2021distributional}. Exact
distributional conformal prediction inverts its accepted rank set rather than
a coarse probability grid. The max-score construction uses its signed
calibration quantile without truncating negative values; the separate
one-sided calibration corrections are truncated below at zero. The rank-floor allocation
provides an allocation-domain endpoint comparison. The candidate allocation grid contains both
allocation-domain endpoints in addition to its interior grid points. All methods share the same
splits and fitted conditional quantiles.

We report
\(\alpha\in\{0.10,0.05,0.02,0.01,0.005\}\). The first two values give
conventional 90 and 95 percent coverage references; the remaining values trace
the transition toward the shrinking-miscoverage regime. The main table uses
\(\alpha=0.05\), where the lower and upper evaluation counts remain stable,
and the figure retains
all five values.

\subsection{Lower and upper miss rates, finiteness, and length}
\label{sec:application-results}

Marginal coverage provides the validity check. The lower and upper miss rates
describe how empirical miscoverage is divided between the tails, finiteness indicates whether the
procedure returns a bounded interval, and relative length is interpreted only
after those properties have been examined. Intervals from the empirical
allocation rule and equal-tail intervals have explicit lower and upper
tail-specific miscoverage levels. For max-score
conformalized quantile regression and exact distributional conformal
prediction, the lower and upper miss rates describe the empirical division of
miscoverage rather than separately guaranteed tail targets. Table \ref{tab:applications} reports means
over 20 paired splits. Parentheses after coverage give Monte Carlo standard
errors. Relative length is the median paired ratio to equal-tail, followed by
its 10th and 90th percentiles.

\FloatBarrier
\begin{table}[!ht]
\caption{Representative application results at \(\alpha=0.05\). Monte Carlo
standard errors are in parentheses; brackets give the 10th and 90th percentiles
of paired length ratios.}
\label{tab:applications}
\centering
\resizebox{\textwidth}{!}{%
\begin{tabular}{llrrrrr}
\toprule
& & Validity & \multicolumn{2}{c}{Miss rates} & Finiteness & Efficiency\\
Data & Method & Coverage & Lower miss & Upper miss & Finite & Rel. length\\
\midrule
freMTPL2 & Empirical allocation & 0.9507 (0.0009) & 0.0002 & 0.0491 & 1.0000 & 0.605 [0.582, 0.634]\\
freMTPL2 & Equal-tail & 0.9500 (0.0007) & 0.0246 & 0.0253 & 1.0000 & 1.000 [1.000, 1.000]\\
freMTPL2 & Max-score CQR & 0.9496 (0.0008) & 0.0209 & 0.0295 & 1.0000 & 0.923 [0.858, 1.000]\\
freMTPL2 & Exact DCP & 0.9498 (0.0010) & 0.0025 & 0.0478 & 0.9987 & 0.620 [0.604, 0.644]\\
\midrule
Wildfire & Empirical allocation & 0.9537 (0.0012) & 0.0000 & 0.0462 & 1.0000 & 0.416 [0.396, 0.435]\\
Wildfire & Equal-tail & 0.9769 (0.0003) & 0.0041 & 0.0190 & 1.0000 & 1.000 [1.000, 1.000]\\
Wildfire & Max-score CQR & 0.9679 (0.0004) & 0.0041 & 0.0280 & 1.0000 & 1.000 [1.000, 1.000]\\
Wildfire & Exact DCP & 0.9527 (0.0005) & 0.0006 & 0.0467 & 0.9998 & 0.447 [0.425, 0.478]\\
\bottomrule
\end{tabular}
}
\end{table}

For claim severity, all four procedures are close to the 0.95 target. For
wildfire size, the empirical allocation rule and exact distributional conformal
prediction are close to target, whereas equal-tail and max-score intervals are
conservative. The empirical allocation rule
has empirical miscoverage concentrated in the upper tail, consistently with the
selected upper-tail allocation fractions. Its median paired length ratios are
0.605 and 0.416. Thus the shorter intervals are reported together with explicit
tail-specific guarantees and finiteness, rather than as the sole comparison
criterion.

\FloatBarrier
\begin{figure}[!ht]
\centering
\includegraphics[width=\textwidth]{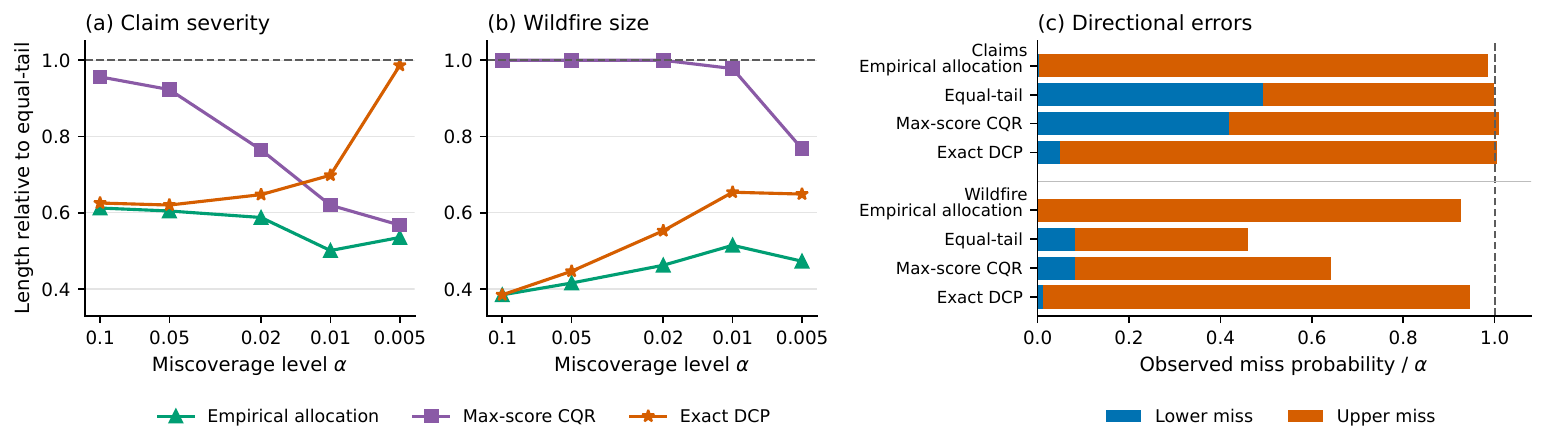}
\caption{Applications: interval length relative to equal-tail for (a) claim
severity and (b) wildfire size, and (c) lower and upper miss rates at
\(\alpha=0.05\).}
\alttext{Three panels summarize two applications. The first two show relative interval length across five miscoverage levels for claim severity and wildfire size, using a shared method legend. The third shows that empirical miscoverage for the empirical allocation rule is concentrated in the upper tail at five percent miscoverage, whereas equal-tail divides empirical miscoverage more evenly.}
\label{fig:applications}
\end{figure}
\FloatBarrier

Across all five miscoverage levels, the empirical allocation rule selects the
rank-floor allocation and remains finite. Its coverage is compatible with each target, and
its median paired length ratio ranges from 0.501 to 0.613 for claim severity
and from 0.385 to 0.515 for wildfire size.
The selected allocation fractions also provide a descriptive indication of the
relative tail behavior represented by the fitted quantiles.
Their concentration at the rank-floor allocation is consistent with the
dominant-upper-tail branch; an interior value would instead be consistent with
both endpoints influencing first-order length.
In panels (a) and (b), the dashed line at one is the equal-tail length
reference. Panel (c), evaluated at \(\alpha=0.05\), decomposes the observed miss
probability into lower and upper parts; its dashed line marks total miss
probability \(\alpha\).

Exact distributional conformal prediction also shortens equal-tail intervals,
but its accepted rank set produces positive infinite-interval rates. This rate
reaches 0.416 for claims at \(\alpha=0.005\), so the finite median length alone
does not summarize its output. Max-score conformalized quantile regression
remains finite but does not separately control the two tail errors. In the
wildfire analysis, its signed calibration quantile is numerically zero. With
many repeated response values and forest quantiles, its displayed median length
therefore matches equal-tail at the three larger miscoverage levels, although the two
calibrations have different lower and upper miss rates. Taken together, coverage,
lower and upper miss rates, finiteness, and length separate the effect of tail
allocation from differences in score construction.

The insurance policies are treated as exchangeable units. The wildfire primary
analysis uses random event splits and targets the corresponding event
distribution. A spatial-cluster-blocked sensitivity in the Supplementary
Material assigns each geographic cluster to one split role. It is a descriptive
check on cluster-blocked random splitting, not a held-out-region analysis or an
additional distribution-free guarantee.

\section{Discussion}
\label{sec:discussion}

Equal-tail allocation is optimal when the two endpoint scales match; conformal
calibration itself does not favor symmetry. In regularly varying tails, the
allocation-length function places the optimum at the boundary when one tail
dominates and in the interior when both tails contribute at first order. The
empirical allocation rule attains either optimum without estimating a tail
index or scale ratio, while separate one-sided calibration preserves marginal
coverage. Under tail homogeneity the regularly varying length relation transfers
over covariates. Groups with opposite dominant tails instead require allocations
that differ across groups because no common fraction is efficient for both.

Two axes explain where this conclusion changes form. The first, tail-quantile
response, distinguishes the multiplicative geometry of regular variation from
the additive geometry of the de Haan class. The second, calibration resolution,
distinguishes diverging tail counts from the bounded-count regime, where
Poisson order-statistic variation and two-sided rank feasibility enter the
first-order limit. The latter boundary identifies where ordinary two-sided
intervals lose finiteness and tail extrapolation becomes relevant.

In implementation, the candidate allocation grid should include both allocation-domain
endpoints; the numerical studies use an interior spacing of 0.005. The selected
allocation fraction is most informative when reported with the lower and upper miss rates
and the finite-interval rate. An endpoint value summarizes one-sided dominance
within the candidate allocation grid, while an interior value indicates that both
endpoints influence length. Neither interpretation changes the marginal nature
of the coverage guarantee. The exact one-sided prediction sets occur at
\(c=0\) and \(c=1\); allocation-domain endpoint selections are their finite
two-sided analogues under the ordinary rank floor.

Uniform extreme conditional quantile theory would replace the high-level
estimation condition in Theorem \ref{thm:transfer} for broader estimator
classes. Heterogeneous tail dominance requires a localized allocation together
with a compatible calibration construction. The functional length process in
the Supplementary Material also suggests simultaneous inference for allocation
curves and local distribution theory for the selected allocation fraction.
The tail-response classification here concerns unbounded upper quantiles. A
bounded-endpoint max-domain requires the distance to the finite endpoint as its
natural scale and leads to a separate allocation problem.

\section*{Supplementary Material}

The Supplementary Material contains complete proofs, finite-rank and
second-order results, full simulation definitions and numerical summaries, and
complete five-level application tables. It also reports the wildfire
cluster-blocked sensitivity. Replication code and data-processing scripts
accompany the supplementary files.

\section*{Data availability}

\begin{sloppypar}
The freMTPL2 data are available through the \texttt{CASdatasets} package
\citep{dutang2026casdatasets}. The FPA FOD-Attributes data are available from
Zenodo under DOI 10.5281/zenodo.8381129
\citep{pourmohamad2023data}.
\end{sloppypar}

\bibliographystyle{plainnat}
\bibliography{references}

\end{document}